\documentclass[3p,times,twocolumn]{elsarticle}

\usepackage{amssymb}

\usepackage[figuresright]{rotating}

\newcommand{\pt}{ p_{\rm t}}

\begin{document}

\begin{frontmatter}



\title{QGP formation time and the large photon v2 puzzle in heavy ion collisions}

\author{Fu-Ming Liu}
\address{Institute of Particle Physics and Key laboratory of Quark \& Lepton
Physics (Ministry of Education), Central China Normal University,
Wuhan, China}


\begin{abstract}

We investigate the large photon $v_2$ puzzle and the two time scales (thermal and chemical equilibrium) in heavy ion collisions. The two-time-scale picture has a weak effect on the transverse moemntum spectrum of direct photons, but a strong effect on the elliptic flow of direct photons.  Thus both the spectrum and the elliptic flow of direct photons may be explained with  hydro evolution constrained with hadron data. In such a picture, a gluon dominant matter appears in heavy ion collisions. This new matter may impact strongly to other fields such as astrophysics and cosmology.
\end{abstract}


\end{frontmatter}


 
 One of the main goals of relativistic heavy ion physics is to study the properties of quark gluon plasma (QGP) which is believed to have existed in the early universe according to the big bang theory.  
Various QGP signatures have been observed in relativistic heavy ion collisions, ie jet quenching and thermal photon emissions, etc. Thermal photons have been observed in heavy ion collisions as the excess to the direct photon emission in cold nuclear collisions, ie in Au+Au collisions~\cite{PHENIX3} at $\sqrt{s_{NN}}$=200~GeV and Pb+Pb collisions~\cite{ALICE2} at $\sqrt{s_{NN}}$=5.5~TeV.

What’s the large photon v2 puzzle in heavy ion physics? In the non-central heavy ion collisions, the elliptic flow $v_2$ of direct photons is measured as large as that of hadrons~\cite{ALICE2,PHENIX2}. It is puzzling, because It is widely believed that the strong emission of photons at early stage should carry a very small $v_2$ while hadrons emitted at the later and cooler stage should carry a larger one.  An even more challenging observation~\cite{PHENIX2} is that the recent measured triangular flow $v_3$ of direct photons seems to be as large as that of hadrons, too. Many efforts have been made by many groups \cite{Linnyk:2013hta,Shen:2013cca,Basar:2012bp,Muller:2013ila} to understand this measured large elliptic flow of direct photons but not succeeded.  We made some progress~\cite{FML2014} on this puzzle with a two-time-scale picture, namely distinguishing thermal and chemical equilibrium time in the way to form a QGP, and extracted the time for QGP formation is about 2~fm/c. In this paper we will explain how this idea was developed and make some explanation to the questions I face so often. 
 
What are thermal equilibrium and chemical equilibrium in heavy ion collisions? 
Thermal equilibrium is an important condition in thermodynamics. Check the energy distribution of constituent particles in local rest frame. If it satisfies the thermal distribution, ie,  Bose-Einstein distribution for Bosons and Fermi-Dirac distribution for Fermions, then the system has reach thermal equilibrium. Note observations are made in lab frame, not in local rest frame.  So for each particle,  the energy and momentum between the two frames are connected with a Lorentz boost with flow velocity, the velocity of collective motion. This flow velocity is important to observe nonzero  elliptic flow of photons in lab frame, because photons are emitted isotropic and elliptic flow vanishes in the local rest frame.

Chemical equilibrium in heavy ion collisions sounds strange because there is neither molecule nor real chemical reaction in heavy ion physics.  A QGP is made of quarks, antiquarks and gluons in both thermal and chemical equilibrium.  Chemical equilibrium requires no change of the net number of quarks, antiquarks or gluons for a given space-time cell. 
 
 What's the energy/momentum distribution for quarks and gluons at the very beginning?  What are the microscopic processes  to realize the QGP formation, or reach the two equilibriums, respectively? 
Normally  the two equilibrium are simply assumed to reach at the same time, the hydro initial time $\tau_0$.  
But from the accelerated projectile nucleus A and target nucleus B, the energy and momentum distribution of quarks and gluons should be obtained from the nuclear parton distribution functions (PDF) based on the global fits to the large set of experiment data, not an easy work. However, kinetics tells partons with big $x$ go to the leading rapidity region. Partons with small $x$ involve the QGP formation, where gluons are much more populated than quarks.

Which of the two equilibriums reaches earlier?
Both elastic processes ie,  $qq\rightarrow qq$, $ qg\rightarrow qg$, $gg\rightarrow gg$ etc and inelastic
processes $gg \rightarrow q \bar q$ and $q \bar q \rightarrow gg$ drive the syetem to local thermal equilibrium. Only the inelastic process $gg \rightarrow q \bar q$ help construct chemical equilibrium. So  thermal equilibrium reaches much faster/earlier  than chemical equilibrium, the same as the chemical common sense, when  no heat exchange with the environment.

How can this two-time-scale picture solve the big photon $v_2$ puzzle?
First we should note that before chemical equilibrium, the system is dominant with gluons and lack of quarks. Main processes to get photons are $q\bar q \rightarrow g \gamma$ and $qg \rightarrow q \gamma$. Processes $gg\rightarrow \gamma +X$ are forbidden when no quark appears.  Photon emission rate is suppressed before QGP formation.  

\begin{figure}
\includegraphics[width=0.45\textwidth]{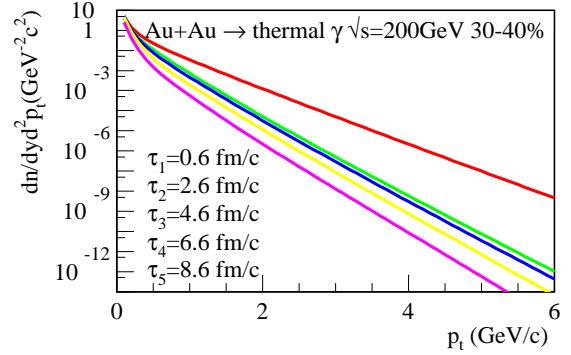}

 \caption{\label{Fig1} (Color Online) Contribution of thermal photons at different time to the transverse momentum spectrum. Curves from up to down are $\tau=0.6, 2.6, 4.6, 6.6$ and 8.6~fm/c, respectively.   This calculation is based Hirano's hydro table for AuAu collisions at $\sqrt{s_{NN}}=200$~GeV  for centrality 30-40\%.   }
\end{figure}

Now we show thermal photons at different time with a normal calculation.  Both  local thermal equilibrium and chemical equilibrium are assumed reached at the hydro initial time, say 0.6~fm/c, as an example with Hirano's hydrodynamics~\cite{Hirano}. This hydro model is well constrained with hadron data.
The $\pt$ spectrum of thermal photons reads \begin{equation}
\frac{dN}{dyd^{2}\pt}=\int d^{4}x\Gamma(E^{*},T)\label{eq:E*a}\end{equation}
 where $\Gamma(E^{*},T)$ is the photon emission rate at temperature
$T$ and $E^{*}=p^{\mu}u_{\mu}$, $p^{\mu}$ is the four-momentum
of a photon in the lab frame and $u_{\mu}$ is the flow velocity. $d^4 x= dxdyd\eta \tau d\tau$ is the space-time cell. If we replace the time integrate $d\tau$ with a unit time interval, we can see how strong the photons are emitted at different time.  An example is shown in Fig.~\ref{Fig1}, for 30-40\%  AuAu collisions at $\sqrt{s_{NN}}=200$~GeV.
 The curves from up to down are $\tau=0.6, 2.6, 4.6, 6.6$ and 8.6~fm/c, respectively.  The early emission is very intense, if no distinguish between the two equilibrium time scales.  

\begin{figure}
\includegraphics[width=0.45\textwidth]{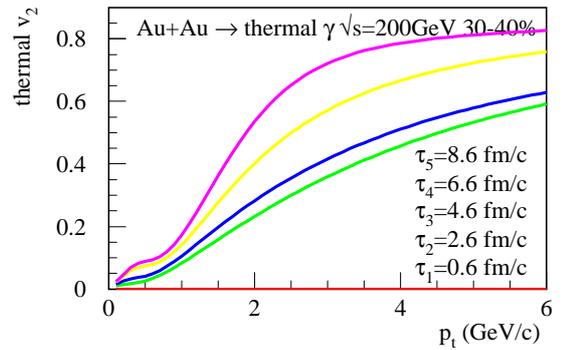}

 \caption{\label{Fig1b} (Color Online) Elliptic flow $v_2$ of thermal photons at different time. Curves from down to up are $\tau=0.6, 2.6, 4.6, 6.6$ and 8.6~fm/c, respectively. This calculation is based Hirano's hydro table for AuAu collisions at $\sqrt{s_{NN}}=200$~GeV  for centrality 30-40\%.  Note at the initial time $\tau=0.6$~fm/c the elliptic flow vanishes because initial flow velocity is zero. }
\end{figure}

We can also check the elliptic flow $v_2$ carried by the photons emitted at different time.
The $\pt$-spectrum of thermal photons emitted at a given time $\tau$ can also be decomposed into harmonics of azimuthal angle $\phi$ and obtain the elliptic flow.
In Fig~\ref{Fig1b} is shown the elliptic flow $v_2$ of thermal photons at different time. Evidently, with the increase of flow velocity (veclocity of collective motion), the later emitted thermal photons carry bigger $v_2$.  Thus if we can suppress the earlier thermal photon emission, we may increase the overall averaged $v_2$ of direct photons and let the measured big $v_2$ explained.
 
\begin{figure}[htb]
\includegraphics[width=0.45\textwidth]{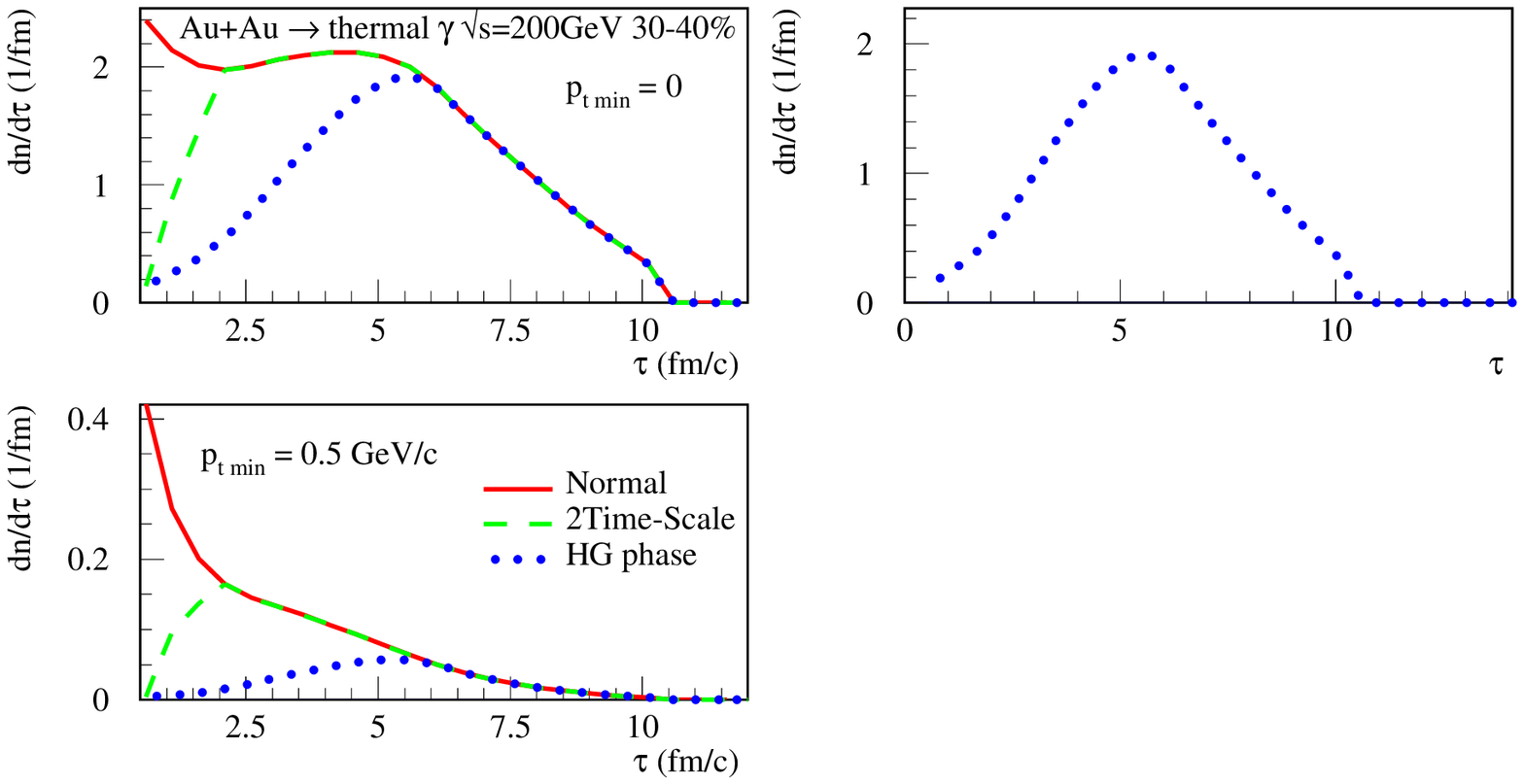}

 \caption{\label{Fig2a} (Color Online)  Number of thermal photons emitted at different time $\tau$, with normal calculation (red lines), two-time-scale calculation (dashed lines) and from hadron phase ( dotted lines). Upper (lower) panel photons are counted with transverse momentum bigger than zero ( 0.5~GeV/c). }
\end{figure}

So the two-time-scale picture can work, because the stage before QGP formation is in fact gluon dominant system and photon emission rate is suppressed. We model the emission rate before QGP formation to do the calculation and noted as "2Time-Scale". The emission rate at a space-time point $x=(\vec r, \tau)$ before QGP formation is assumed as
\begin{equation}
\Gamma(x)=(\frac{\tau - \tau_{\rm QGP}}  {\tau_0-\tau_{\rm QGP}})^2 \Gamma_{QGP} (T (x))
\label{eq:rate}\end{equation}
where $\tau_0=0.6$~fm/c is the beginning time for local thermal equilibrium and hydrodynamic expansion, $\tau_{\rm QGP}=2.1$~fm/c is the beginning time for chemical equilibrium and photon emission with full QGP rate, AMY rate~\cite{AMY2001} $\Gamma_{QGP}$ as  a function of temperature $T$.  Here we don't extract the QGP formation $\tau_{\rm QGP}$ but directly take the value from previous paper~\cite{FML2014}. 
We also show the normal calculation and noted as "Normal" where QGP is assumed to form at the initial time 0.6~fm/c. Photon emission in hadonic gas phase remains not modified, and shown as "HG phase".

In Fig~\ref{Fig2a} is shown the number of thermal photons emitted at different time $\tau$, with normal calculation (red lines), two-time-scale calculation (dashed lines) and from hadron phase ( dotted lines).
 Photons are massless.  Photon number depends strongly how small energy photons are counted. In upper (lower) panel,  photons are counted with transverse momentum bigger than zero (0.5~GeV/c). 
Thus Fig~\ref{Fig2a} shows us clearly only photon emission from partonic phase before QGP formation is modified in this two-time-scale picture, and how much is modified. 
Since the early photon emission is suppressed by the delayed QGP formation,  the overall elliptic flow will get bigger and closer to the value carried by  later emitted photons.

\begin{figure}[t]
\includegraphics[width=0.45\textwidth]{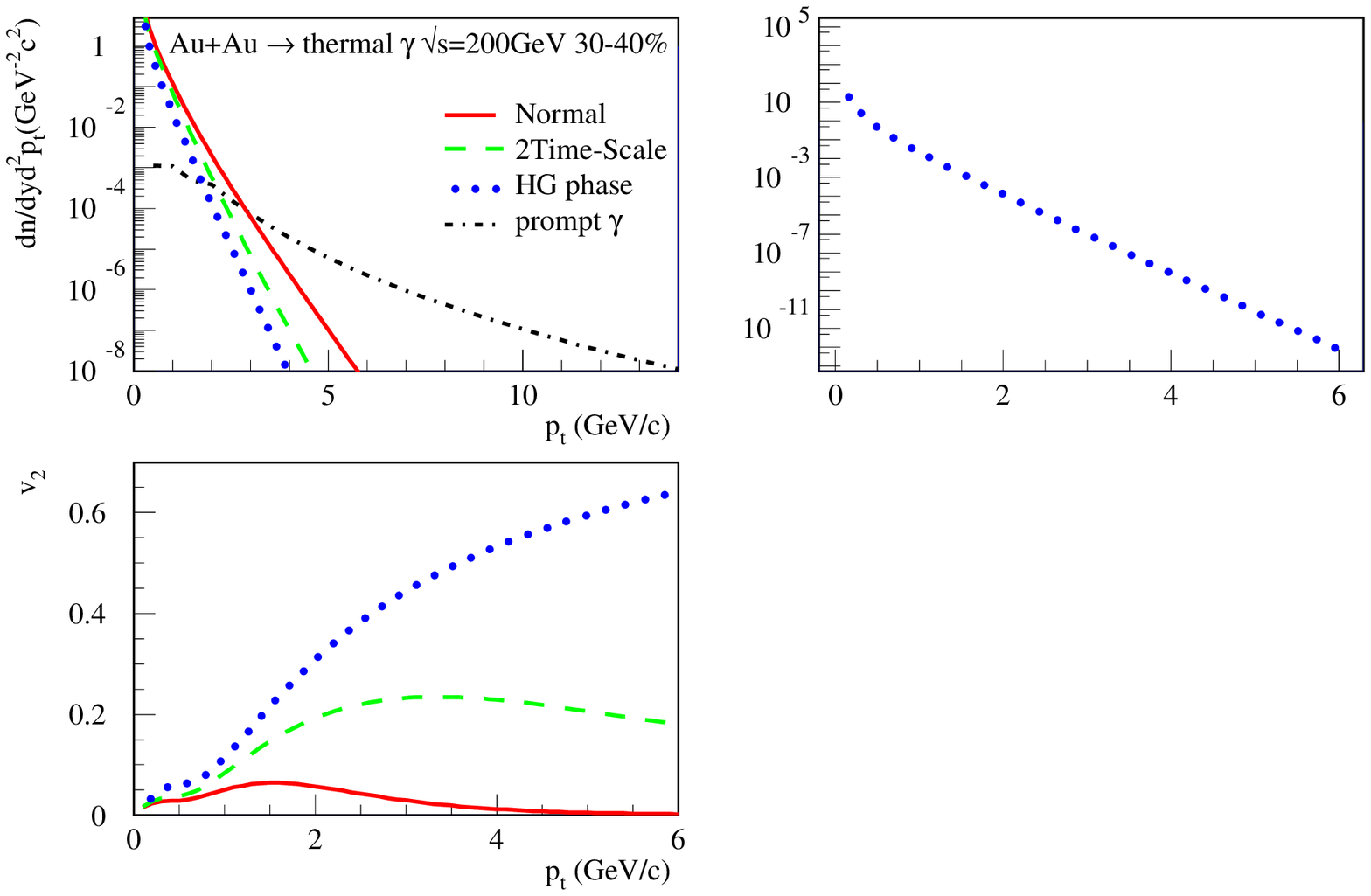}

 \caption{\label{Fig2b} (Color Online)  The $\pt$ spectrum (upper panel) and the elliptic flow (lower panel)
 of thermal photons from normal calculation (red solid lines) , two-time-scale calculation (green dashed lines) and from hadronic phase (blue dotted lines). The $\pt$ spectrum of prompt photons is plotted as dashed dotted line in upper panel.  For AuAu collisions at $\sqrt{s_{NN}}=200$~GeV  with centrality 30-40\%.  }
\end{figure}

Let's check the time-integrated results. In Fig.~\ref{Fig2b} 
is shown the $\pt$ spectrum (upper panel) and elliptic flow (lower panel) of thermal photons from normal calculation (red solid lines) , two-time-scale calculation (green dashed lines). Photon number reduces from the normal calculation to the two-time-scale calculation, however elliptic flow increases. Photons from hadronic phase (blue dotted lines) is an important contribution  at very low $\pt$. And the elliptic flow is large due to the later emission.
As a reference, the contribution of prompt photons is shown as dashed dotted line in upper panel. It is calculated to the next-to-leading order contribution in cold nuclear collisions.
One can see the prompt photons are the main contribution when $\pt $ is bigger than  3~GeV/c  (2~GeV/c) in normal (two-time-scale) calculation. The prompt photons are produced before hydro expansion, and carries vanishing elliptic flow. They will suppress the elliptic flow of thermal photons. 
 
\begin{figure}[htb]
\includegraphics[width=0.45\textwidth]{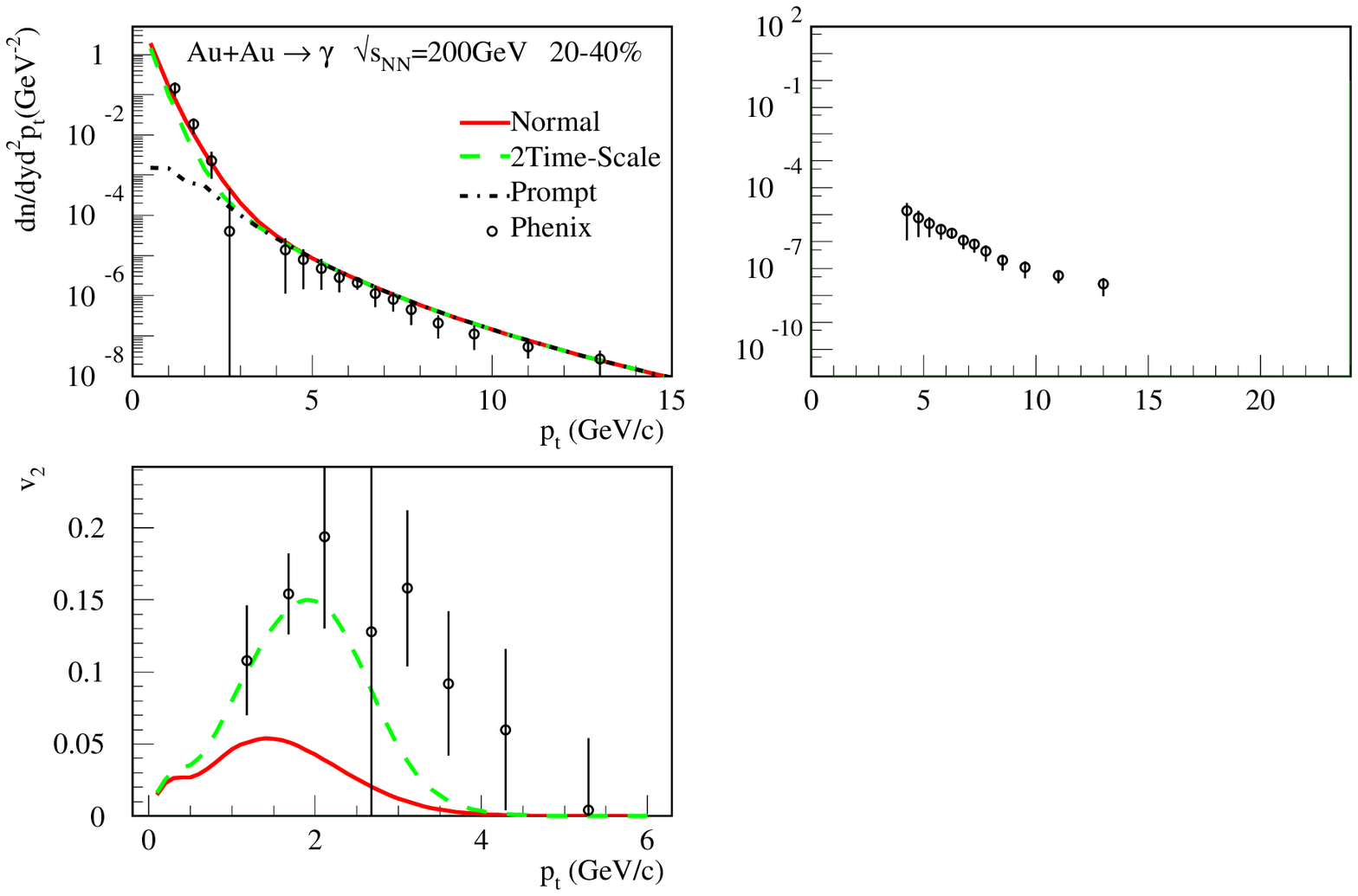}
 \caption{\label{Fig3} (Color Online)   The $\pt$ spectrum (upper panel)  and elliptic flow $v_2$ (lower panel) of direct photons(upper panels) from AuAu collisions at $\sqrt{s_{NN}}=200$~GeV  for centrality  20-40\% ). Normal calculation is shown as red solid lines and  the two-time-scale  as green dashed lines. Prompt photons are plotted as dashed dotted line. Data points from PHENIX. }
\end{figure}

How much will the two-time-scale picture  effect the $\pt$ spectrum of direct photons? 
In Fig.~\ref{Fig3} upper panel  the $\pt$ spectrum of direct photons (thermal + prompt photons) is compared with the experimental data (open cycles). It might be surprised to see both the normal (red solid line) and the two-time-scale (green dashed line) coincidence with the measured spectrum. However, we may notice: 
At high $\pt$, prompt photons are the dominant contribution to direct photons. This contribution is the same in the two calculations, normal and two-time-scale.   
The two calculations of thermal photons only differ before QGP formation. The resulted $\pt$ spectrum as shown in Fig.~\ref{Fig2b} (upper panel) differ very little above the prompt photon line. 
That's the two-time-scale picture has a very weak effect on the $\pt$ spectrum of direct photons.

However, the effect of the two-time-scale picture on the elliptic flow is very strong, as shown in Fig.~\ref{Fig3} lower panel. The elliptic flow of direct photons with normal calculation (red solid line) is much lower than the experimental data points (empty cycles), because of the strong emission at the early time. 
With the two-time-scale calculation (green dashed line)  the early emission is suppressed, so the overall elliptic flow of thermal photons increases quite a lot and goes closer to data points.

The two-time-scale picture can explain both the $\pt$ spectrum and the elliptic flow of direct photons.  But does the hydrodynamic evolution is still hadron data constrained? The answer is yes. Dynamical evolution is related force and mass.  Before QGP formation, the pressure (energy density gradiant) is the same as normal calculation. It drives the  mass of the bulk to move collectively outward. This force is blind to quarks and gluons. This is similar to the end of movie when people move collectively out of the cinema. The driving force is people density gradient, which is blind to men and women.  Finally the QGP is formed, at a moment earlier  than hadron freeze-out. So bulk hadrons can not tell us the precise time of  QGP formation.

Finally, this two-time-scale picture reminds us the creation of a new kind of matter in heavy ion collisions, a gluon dominant plasma, or shortly glasma. If it exists in the universe, it will show mass heavy (gravity) but shine less brightly. The universe may offer us a rich experimental lab with much more extreme conditions than we can make in the planet. Please do not limit our mind with what we see ordinarily.

This work was supported by the Natural Science Foundation of China
under Project No.11275081 and by the Program for New Century Excellent
Talents in University (NCET).




\nocite{*}
\bibliographystyle{elsarticle-num}
\bibliography{martin}



\end{document}